%% file: nfkb.tex
\documentclass[twocolumn]{article}
\usepackage{graphicx,color}
\usepackage{epsfig}

\begin{document} 
\bibliographystyle{pnas1}
                                                                                                                                                                    
\title{\textbf{\LARGE Spiky oscillations in NF-$\kappa$B signalling}}
\author
{Sandeep Krishna,$^{*}$ Mogens H. Jensen, Kim Sneppen\\
\\
\normalsize{Niels Bohr Institute, Blegdamsvej 17, 2100 Copenhagen \O, Denmark.}\\
\\
\normalsize{$^*$To whom correspondence should be addressed; E-mail:  sandeep@nbi.dk.}
}
\date{}

\maketitle 

{\bf
The NF-$\kappa$B signalling system is involved in a variety of cellular
processes including immune response,
inflammation, and apoptosis.
Recent experiments
have found oscillations in the nuclear-cytoplasmic translocation
of the NF-$\kappa$B transcription factor 
[Hoffmann {\it et al.} (2002) {\it Science} {\bf 298}, 1241;
Nelson {\it et al.} (2004) {\it Science} {\bf 306}, 704.]
How the cell uses the oscillations to differentiate input conditions
and send specific signals to downstream genes is an open problem.
We shed light on this issue by examining the
small core network driving the oscillations, which, we show,
is designed to produce periodic spikes in nuclear NF-$\kappa$B
concentration. The oscillations can be used to regulate downstream genes in
a variety of ways. In particular, we show that 
genes to whose operator sites NF-$\kappa$B binds and dissociates fast
can respond very sensitively to changes in the input signal, with
effective Hill coefficients in excess of 20.}

NF-$\kappa$B is a family of dimeric transcription factors which participate in the
regulation of a number of cellular 
processes including immune response, inflammation and apoptosis
\cite{LDHSG,LBLNK,GhoshKarin,GMK}.
Extensive experiments using electrophoretic mobility shift assay
and single-cell flourescence imaging
have found oscillations in the nuclear-cytoplasmic translocation
of the NF-$\kappa$B transcription factor in mammalian cells \cite{HLSB,Nelsonetal},  
with a time period of the order of hours.
NF-$\kappa$B can be activated by a number of external stimuli \cite{Pahl}
including bacteria, viruses and various stresses and proteins (e.g., tumor necrosis factor-$\alpha$, TNF-$\alpha$, 
which was the signal used in refs.
\cite{HLSB,Nelsonetal}).
In response to these signals it targets over 150 genes
including many chemokines, immunoreceptors, stress reponse genes, as 
well as acute phase inflammation
response proteins \cite{Pahl}.
Experiments show that NF-$\kappa$B does not regulate all
its downstream genes in the same way. For example, the
chemokine gene RANTES turns on much later than 
another chemokine IP-10 after TNF-$\alpha$ activation \cite{HLSB}.
Thus, the two main questions raised by the dynamics of the NF-$\kappa$B
system are: how does the network of interactions produce
oscillations, and how does the cell use
the oscillations to differentiate input conditions
and send specific signals to downstream genes?
In this paper, we elucidate 
the small core network driving the oscillations and show that it
is designed to produce periodic spikes in nuclear NF-$\kappa$B
concentration. We show that the spiky oscillations are extremely robust 
to variation of parameters. We further argue that the spikiness
is associated with an increased sensitivity
of the system that could be used 
for differentially regulating downstream genes. 

\section*
{Extracting the Core Feedback Loop}
\begin{figure}[t]
{\bf A}\\
\centerline{\epsfig{height=5cm,file=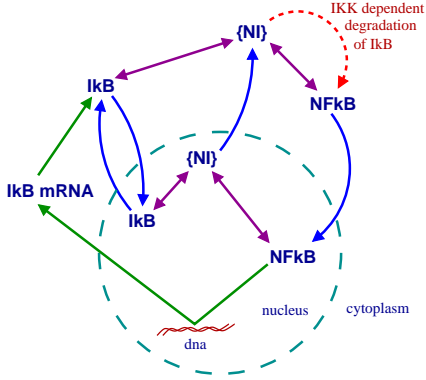}}
{\bf B}\\
\centerline{\epsfig{height=5cm,file=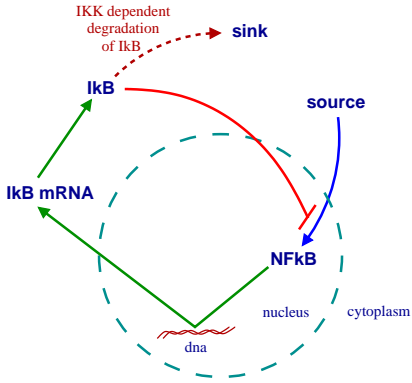}}
\caption{
\label{schematic}
Schematic diagram of interactions in the NF-$\kappa$B signalling system.
{\bf A}. 7-variable model.
{\bf B}. 3-variable model.
}
\end{figure}

Hoffman {\it et al.} have constructed
a long list of chemical reactions
between 26 different molecules in the NF-$\kappa$B system,
including reaction constants \cite{HLSB}.
We reduced this system
to the core feedback loop (Fig. \ref{schematic}B) generating oscillations.
The reduction 
was done in three steps: the first, removing molecules which 
have no feedback from NF-$\kappa$B and 
deleting slow reactions where faster alternate pathways exist 
(e.g., export of nuclear NF-$\kappa$B) 
resulted in a 7-variable model. 

The interactions in
this model are schematically displayed in Fig. \ref{schematic}A.
It consists of cytoplasmic and nuclear NF-$\kappa$B,
its inhibitor, 
I$\kappa$B, and I$\kappa$B kinase (IKK) which phosphorylates the inhibitor,
leading to its degradation.
The inhibitor forms a complex with NF-$\kappa$B which, in the cytoplasm,
prevents its transport into the nucleus. Only free nuclear NF-$\kappa$B is imported 
into the nucleus. In contrast, from inside the nucleus,
only the complex can be exported, not the free NF-$\kappa$B.
I$\kappa$B is known to occur in several isoforms.
Cells containing only the I$\kappa$B$\alpha$ isoform show sustained
oscillations, while cells with only the I$\kappa$B$\beta$ or $\epsilon$
isoforms do not show oscillations. Wild type cells, with all three isoforms,
typically exhibit damped oscillations \cite{HLSB}.
The difference between these isoforms is that only I$\kappa$B$\alpha$
is activated by NF-$\kappa$B \cite{SGBG,SFLNB}. In contrast 
I$\kappa$B$\beta,\epsilon$ are produced at a rate
independent of NF-$\kappa$B and so lie outside the feedback loop
(see supporting text for the equations
governing the dynamics of the 7-variable model.)

Coarse graining over fast chemical reactions involving 
complex formation reduced this system
to 4 variables.
Finally, based on numerical observations, we found we could 
effectively eliminate 
nuclear I$\kappa$B, giving the model in Fig. \ref{schematic}B.
More details of the reduction process are given 
in supporting text.

\section*{Results}
\subsection*{3-Variable Model of NF-$\kappa$B Oscillations}
The core feedback loop,
we find, 
consists
of only three constituents (Fig. 1B):
nuclear NF-$\kappa$B ($N_n$), cytoplasmic I$\kappa$B ($I$)
and I$\kappa$B mRNA ($I_m$). 
NF-$\kappa$B dimers activate production of
I$\kappa$B mRNA 
which translated to I$\kappa$B
inhibits nuclear 
NF-$\kappa$B production, completing the feedback loop. 

The dynamics of the system in Fig. \ref{schematic}B is captured by
three coupled ordinary differential equations:
\begin{eqnarray}
\frac{dN_n}{dt}&=&A\frac{(1-N_n)}{\epsilon+I}-B\frac{IN_n}{\delta+N_n},\\
\frac{dI_m}{dt}&=&N_n^2-I_m,\\
\frac{dI~~}{dt}&=&I_m-C\frac{(1-N_n)I}{\epsilon+I}.
\end{eqnarray}
For ease of analysis we have rescaled all variables
to be dimensionless (the original equations 
and the rescaling process are described in supporting text.)
$A, B, C, \delta, \epsilon$ are dimensionless parameters dependent on the reaction constants
(see supporting text).
The external signal is supplied by IKK that enters the equations
through the parameter, $C$, which is proportional to IKK concentration.
The model has no spatial degrees of freedom. It has been suggested that
flow induced stresses can trigger NF-$\kappa$B mediated gene activity \cite{Gangulietal}
but here we assume that each compartment, the cytoplasm and nucleus, is 
well-mixed.

The first term in equation (1a) represents the import of free cytoplasmic 
NF-$\kappa$B (whose concentration is $1-N_n$) into the nucleus. This is 
hindered by the presence of cytoplasmic I$\kappa$B which sequesters
NF-$\kappa$B in the cytoplasm. Parameter $A$ is proportional to the
NF-$\kappa$B nuclear import rate. The second term in (1a) derives from the export of
NF-$\kappa$B from the nucleus via the NF-$\kappa$B--I$\kappa$B complex,
which is why the term also depends on $I$. $B$ is proportional to
I$\kappa$B nuclear import rate. $\delta$ sets the concentration at which
half the nuclear I$\kappa$B is complexed to NF-$\kappa$B and it depends
both on the rates of association and dissociation of the complex as well
as the export rate.

The first term in equation (1b), the rate of production of I$\kappa$B mRNA,
contains the square of $N_n$ because the production is activated by
NF-$\kappa$B dimers\footnote{Ref. \cite{Nelsonresponse} argues that making
this term linear in $N_n$ is better.  
We have checked that this change does not alter our
conclusions (see supporting text.)}. 
The second term is the degradation of the mRNA whose rate,
in these rescaled equations, sets the overall timescale.
It is easy to modify this equation to deal with the $\beta$ and $\epsilon$
isoforms of I$\kappa$B simply by adding a constant for their NF-$\kappa$B-independent
rate of production.

Equation (1c) has one term for the production of cytoplasmic I$\kappa$B from
its mRNA and a second for its degradation due to the presence of IKK.
This degradation is proportional to the concentration of the NF-$\kappa$B--I$\kappa$B
complex, which depends on both $I$ and $(1-N_n)$, the concentration of
cytoplasmic NF-$\kappa$B. $\epsilon$ sets the concentration at which half of
the cytoplasmic NF-$\kappa$B is in the complex. $C$ is proportional to the rate of degradation
and to the IKK concentration.

\begin{figure}[t]
\centerline{\epsfig{height=5cm,file=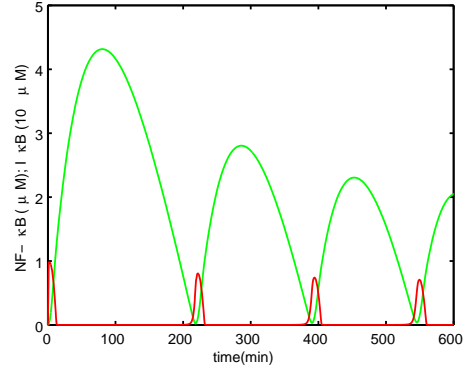}}
\caption{
\label{osc}
Sustained oscillations of nuclear NF-$\kappa$B ($N_n$), red, and
cytoplasmic I$\kappa$B, green, for $A=0.007$, $B=954.5$, $C=0.035$, $\delta=0.029$
and $\epsilon=2\times 10^{-5}$ (using parameter values
taken from \cite{HLSB}, see supporting text). The shape, phase and period are remarkably similar
to the experimental plot in Fig. 3B, left panel, in ref. \cite{Nelsonetal}.
}
\end{figure}

Fig. \ref{osc} shows a plot of nuclear NF-$\kappa$B and
cytoplasmic I$\kappa$B concentrations obtained in simulations, using parameter
values from ref. \cite{HLSB}.
Our model predicts the following experimentally observed facts \cite{HLSB,Nelsonetal}:
i) sustained oscillations in cells with only the $\alpha$ isoform of I$\kappa$B,
ii) damped oscillations in wild type cells which include other isoforms of I$\kappa$B, 
iii) time period of the order of hours, 
iv) spikiness of nuclear NF-$\kappa$B 
and asymmetry of 
cytoplasmic I$\kappa$B oscillations,
v) phase difference between NF-$\kappa$B and I$\kappa$B,
vi) lower frequency upon increased transcription of I$\kappa$B.\\

\subsection*{Saturated Degradation of I$\kappa$B is Crucial for Oscillations}
A key element in our 3-variable model 
is the saturated degradation of cytoplasmic I$\kappa$B 
in the presence of IKK (second term in eq. 1c)
due to the
Michaelis-Menten complex formation between NF-$\kappa$B and I$\kappa$B --
a complex needed for IKK triggered degradation of I$\kappa$B.
The same complex inhibits nuclear
NF-$\kappa$B production because only free cytoplasmic NF-$\kappa$B is
imported into the nucleus.

\begin{figure}[ht]
\centerline{\epsfig{height=5cm,file=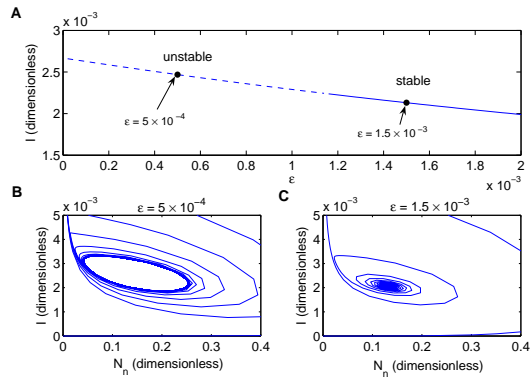}}
\caption{
\label{epsstability}
Importance of saturated degradation of $I$.
{\bf A.} Steady state solution of equation (1) for $I$ as a function of the parameter $\epsilon$.
Dashed line shows the region where the fixed point is unstable and the solid line
shows the region where it is stable 
(see supporting text).
Black dots mark the parameter values used in (B) and (C).
The crossover from unstable to stable occurs where $\epsilon$ becomes comparable to the steady
state value of $I$, i.e., where the degradation of $I$ stops being saturated.
{\bf B.} Trajectory of the system in the $I$-$N_n$ phase plane, for $\epsilon=5\times 10^{-4}$, which converges to
a stable limit cycle. {\bf C.} Trajectory of the system, for $\epsilon=1.5\times 10^{-3}$, which converges
to a stable fixed point.
}
\end{figure}

A stability analysis of the system shows
the importance of the saturated degradation
for oscillations. In general, the system has a single fixed point where
all concentrations are unchanging in time.
For small values of $\epsilon$, corresponding to strong
saturation of the degradation, this fixed point is unstable and the system
goes into a periodic cycle. As $\epsilon$ is increased
the fixed point becomes stable and the oscillations disappear (Fig. \ref{epsstability}).
This happens when the value of $\epsilon$ becomes comparable to the steady state
value of $I$, which is precisely when the degradation rate stops being saturated.
The saturation is crucial for oscillations because it puts an upper limit
to the degradation rate, allowing I$\kappa$B to accumulate and stay around longer
than with the more usual $I$-proportional degradation rate.
This effectively introduces a time delay into the feedback loop.
Negative feedback with time delay is known to easily produce oscillations,
and this mechanism has been used to model the p53 \cite{TSJ} and 
Hes oscillations \cite{JST}.
Here, instead of being put in by hand, the time delay arises more naturally
through the dynamics of the system.

Interestingly, the NF-$\kappa$B core in Fig. \ref{schematic}B is similar
to an early model
showing oscillations by negative feedback \cite{BPM} which
introduced precisely the same kind of saturated degradation to 
mitigate the unreasonably large Hill coefficient in
an even earlier model of Goodwin \cite{Goodwin}.
The importance of this mechanism has also been recognized by Goldbeter who has 
used it in models of various cellular oscillations,
e.g., the cell cycle  \cite{Goldbeter_cellcycle},
development in myxobacteria \cite{Goldbeter_development}, yeast stress response \cite{Goldbeter_yeaststress}
and the mammalian circadian clock \cite{Goldbeter_circadian}.
Saturated degradation has also been implicated in models of calcium oscillations 
in cells \cite{RBSSZE,Goldbeter_calcium}. We further note the p53 system, which also 
shows oscillations, contains the degradation of p53 via the formation of a complex 
with Mdm2 \cite{VLL}, which could result in saturated degradation.
This suggests that saturated degradation might be a very
general mechanism, easily implemented by complex formation and used by
cellular processes to introduce time delays where necessary.

\subsection*{Spiky Oscillations and Control of Downstream Genes}
\begin{figure}[hb]
\centerline{\epsfig{height=5cm,file=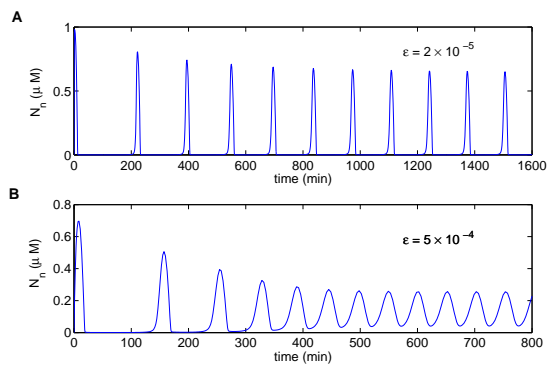}}
\caption{
\label{spikyandsoft}
Spiky and soft oscillations.
Plots of nuclear NF-$\kappa$B concentration as a function of time
from equation (1) with $A$,$B$,$C$ and $\delta$ the same as in Fig. \ref{osc}.
{\bf A.} Spiky oscillations with $\epsilon=2\times 10^{-5}$, the same as in Fig. \ref{osc}.
{\bf B.} Soft oscillations with $\epsilon=5\times 10^{-4}$.
}
\end{figure}

{\bf Robust spiky oscillations.} 
One feature of our model
is that it can produce sharp spikes in nuclear NF-$\kappa$B. 
This is unusual for oscillations driven by negative feedback \cite{Goldbeter_review}. 
We quantify spikiness using the following measure:
$Z=({\rm max}(N_n)-{\rm min}(N_n))/{\rm mean}(N_n)$.
Oscillations with $Z>2$ we term spiky, and oscillations with
$Z<2$, soft.
Fig. \ref{spikyandsoft} shows an example of each type of oscillation, generated
from the 3-variable model using different parameter values.\\

\begin{figure}[p]
{\bf A}\\
\centerline{\epsfig{height=7cm,file=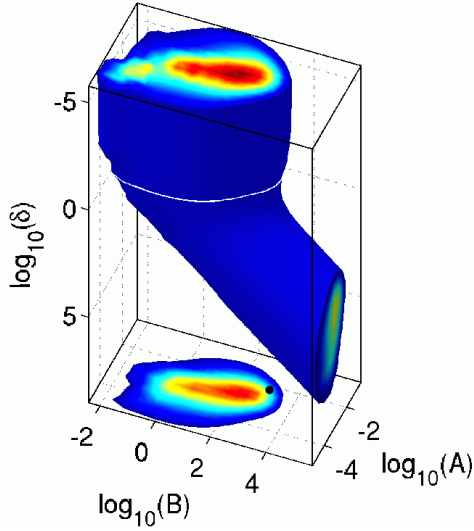}}
{\bf B}\\
\centerline{\epsfig{height=7cm,file=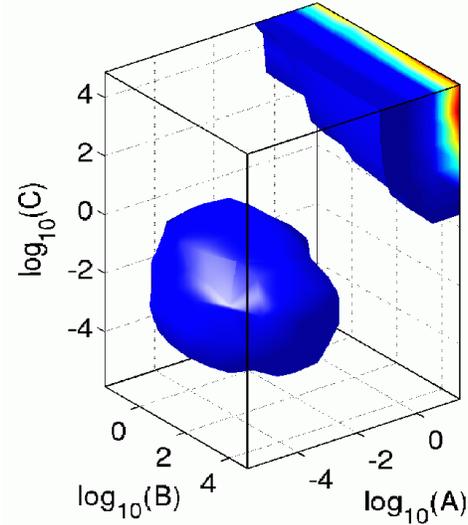}}
\caption{
\label{robustness}
Robustness of spiky oscillations. 
The 3d volume is the region of spiky oscillations ($Z>2$, defined in main text); colour gradient towards red shows increasing values
of $Z$.
{\bf A.} The robustness for parameters $A, B, \delta$. The 2d plot shows the horizontal slice (white line)
through the 3d volume at $\delta=0.029$, the value used in Fig. \ref{osc}.
The black dot corresponds to values of $A,B$ used in Fig. \ref{osc}.
{\bf B.} A similar plot for parameters $A, B, C$.}
\end{figure}

The spikiness is extremely robust to variation
of parameters as shown in Fig. \ref{robustness}.
Note the scales are logarithmic: parameters can be varied
by several decades without going out of the oscillatory regime.
The enormous robustness suggests that the spikiness of the oscillations is
an important element in the design of NF-$\kappa$B signalling
and may be essential for its proper functioning as a transcription factor.
We substantiate this idea in the subsequent sections.\\

\begin{figure}[ht]
{\bf A\hfill B\hfill ~}\\
\epsfig{height=3.5cm,file=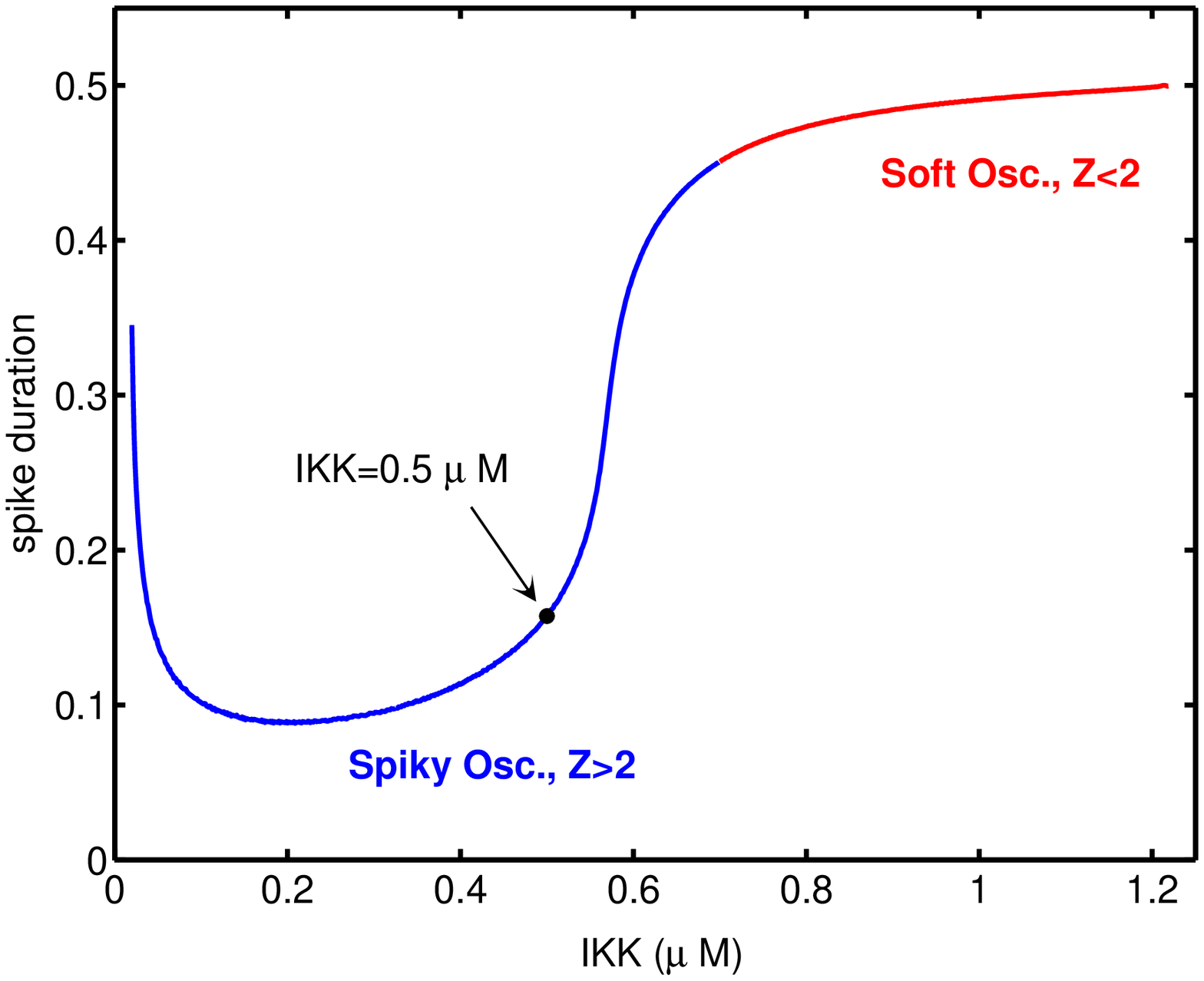}
\epsfig{height=3.5cm,file=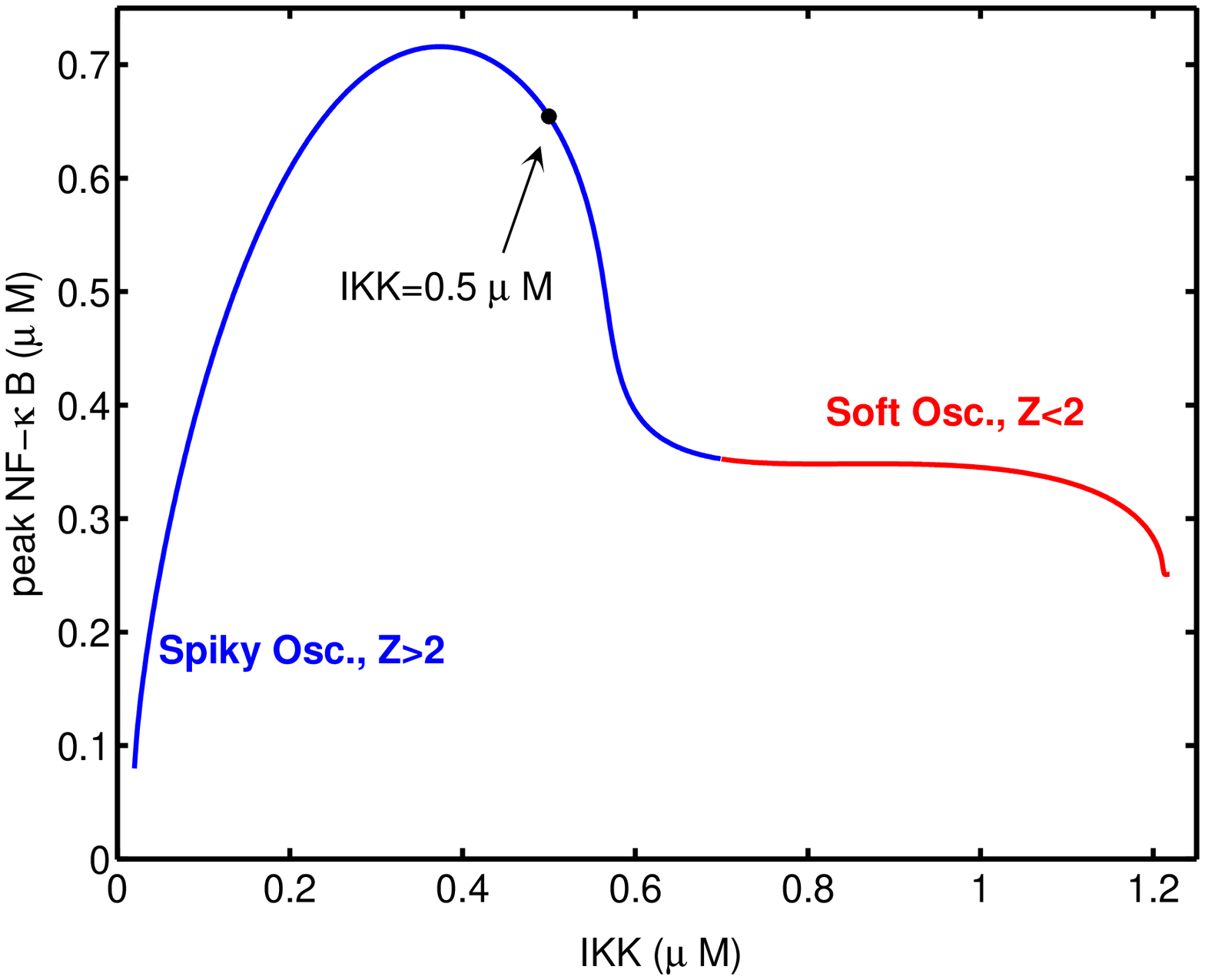}
\caption{\label{IKKsensitivity}
Sensitivity to IKK.
{\bf A.} Spike duration, the fraction of time $N_n$ spends above its mean value, as a function of IKK concentration. The black dot shows the IKK value
used in Fig. \ref{osc}. Blue and red signify, respectively, regions of spiky 
and soft oscillations. Notice the sharp response just before the transition to soft oscillations.
{\bf B.} Spike peak, the maximum concentration of nuclear NF-$\kappa$B, as a function of IKK concentration. The black dot shows the IKK value
used in Fig. \ref{osc}. Blue and red signify, respectively, regions of spiky 
and soft oscillations. Notice the sharp response just before the transition to soft oscillations.
}
\end{figure}

\noindent
{\bf Sensitivity to IKK is high for spiky oscillations.}
Since IKK is the external signal to which the system responds,
we begin by comparing the sensitivity of spiky and soft oscillations
to changes in IKK concentration.
We consider two quantities: the spike duration, defined as the amount of time
$N_n$ spends above its average value, and the spike peak, defined as the
maximum nuclear NF-$\kappa$B concentration during each cycle of oscillations.
Fig. \ref{IKKsensitivity}A shows how the spike duration depends on IKK concentration.
The sensitivity of the spike duration is very high in certain regions of
spiky oscillations.
It is especially large near the transition to soft oscillations.
A similar sensitivity is seen in the peak NF-$\kappa$B concentration (Fig. \ref{IKKsensitivity}B).
Thus, the spike duration and peak
are much more sensitive to (and therefore easier to regulate by)
IKK for spiky
than for soft oscillations.\\

\begin{figure}[t]
\epsfig{height=5cm,file=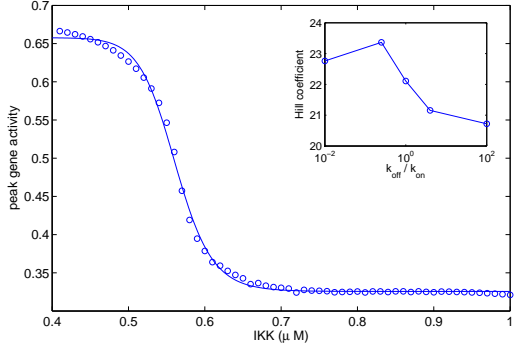}
\caption{\label{eqlgene}
Equilibrium binding of NF-$\kappa$B to a downstream gene.
The plot shows the peak gene activity as a function of IKK concentration (open circles).
The data has been fitted by a sigmoidal function of the form $1/(1+(x/x_0)^h)$. The least squares fit (solid line)
gives an effective Hill coefficient $h=23.4$. 
Inset: effective Hill coefficient obtained by similar fitting for different $k_{off}/k_{on}$ ratios.
}
\end{figure}

{\bf Large Hill coefficients and regulation of downstream genes.}
It is possible for genes regulated by NF-$\kappa$B to 
inherit this sensitivity in the form of a high effective Hill
coefficient. 
Consider a gene which has an operator site at which NF-$\kappa$B dimers
can bind and activate the gene:
$$ G + 2N \mathop{\rightleftharpoons}_{k_{off}}^{k_{on}} G^*.$$
To begin with, we assume that the binding of NF-$\kappa$B to the operator
is in equilibrium, i.e., $k_{on}$ and $k_{off}$ are much larger than
the rates of all other processes in the NF-$\kappa$B system. In that case
the gene activity, $G^*$, will follow the NF-$\kappa$B concentration:
$$G^*=\frac{N_n^2}{k_{off}/k_{on}+N_n^2}.$$
Fig. \ref{eqlgene} shows the peak gene activity as a function of IKK 
concentration. The effective Hill coefficient of this response curve is
over 20, much larger than the values obtained by typical ways of introducing
cooperativity in gene regulation \cite{HF,GK}. 
As the inset shows, the effective Hill coefficient
remains above 20 for a large range of values of the ratio $k_{off}/k_{on}$, i.e.,
genes controlled in this way show a very high sensitivity to the input signal.

\begin{figure}[ht]
\centerline{\epsfig{height=5cm,file=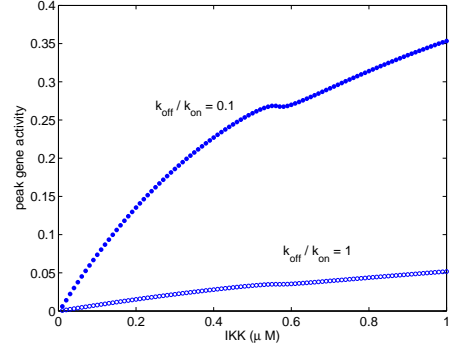}}
\caption{\label{koffsmall}
Nonequilibrium binding of NF-$\kappa$B to a downstream gene.
The plot shows the peak gene activity as a function of IKK concentration,
with $k_{off}=0.001 min^{-1}$ kept fixed while $k_{on}=0.01 \mu M^{-1} min^{-1}$ for filled circles and 
$0.001 \mu M^{-1} min^{-1}$ for open circles. 
}
\end{figure}

When $k_{on}$ and $k_{off}$ become
comparable to other rates in the system the binding of NF-$\kappa$B
to the operator remains out of equilibrium. 
When $k_{off}$ is small enough, the gene activity does not have enough time
to decay completely between spikes of NF-$\kappa$B. Fig. \ref{koffsmall} shows
the peak activity as a function of IKK. In contrast to the equilibrium
case, here the response is linear at best. Note also that the peak gene activity
increases with IKK in contrast to the equilibirum case where it decreases.
Thus, the same oscillations
are capable of regulating genes very differently, depending on their
$k_{on,off}$ values which are determined
by their operator sites.

\begin{figure}[t]
\centerline{\epsfig{height=5cm,file=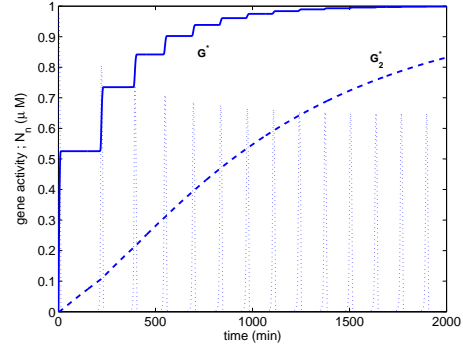}}
\caption{\label{koffzero}
Strongly nonequilibrium binding of NF-$\kappa$B to a downstream gene.
The plot shows nuclear NF-$\kappa$B concentration (dotted line), the
activity of a gene, $G^*$ (solid line), directly activated by NF-$\kappa$B dimers
with $k_{on}=0.1 \mu M^{-2} min^{-1}$, $k_{off}=0$, and the activity of a second
gene, $G_2^*$ (dashed line), that is 
activated by the protein product of $G$ ($k_t=0.01 min^{-1} $,
$\gamma_p=1 min^{-1}$, see main text).
}
\end{figure}

In the extreme case where 
$k_{off}$ is negligible, each spike of nuclear NF-$\kappa$B
contributes to increase the
gene activity until it saturates to unity as shown by the
solid line in Fig. \ref{koffzero}.
If this gene controls the activity of a second one in a cascade:
$$G^*\mathop{\longrightarrow}^{k_t}G^* + P,$$ 
$$P\mathop{\longrightarrow}^{\gamma_p} \phi,$$ 
$$G_2 + P \mathop{\rightleftharpoons}^{k_{on}}_{k_{off}} G_2^*,$$
then the latter will turn on later (dashed line in Fig. \ref{koffzero})
with a time delay that depends on the
timescales of transcription, translation and promoter activation.
This is reminiscent of the experiments of ref. \cite{HLSB} which show the 
gene IP-10 turning on quickly after the introduction of IKK, while the
gene RANTES turns on after a delay.

\section*{Discussion}
What functional role, if any, do the oscillations in the NF-$\kappa$B system
play? There have been several suggestions: downstream gene networks are perhaps
regulated by the frequency of the oscillations, or the oscillations could be a by-product
of rapid attenuation of NF-$\kappa$B,
or they might be used to make multiple evaluations of the
input signal \cite{Lahav,TE}. 
Barken {\it et al.}
\cite{BWKCHL}
warn against overemphasising the physiological role of oscillations.
Our approach to tackling this question has been to construct a
reduced 3-variable model 
which, despite its simplicity, captures many characteristic features
of the system. 
The simplicity of the model allows us to fully explore and understand
the range of dynamical behaviour it exhibits. In particular, we have
shown that it is capable of both spiky and soft oscillations, and
that the spiky oscillations are extremely robust to variation of
parameters.

The activity of genes downstream of NF-$\kappa$B depends upon
the amount of time for which NF-$\kappa$B is present inside
the nucleus in sufficiently large concentrations to dimerize and
bind to those genes' operator sites.
It seems reasonable to assume that NF-$\kappa$B could 
signal different downstream genes, simply by regulating the
amount and exposure time to IKK, provided the signalling system
is sufficiently sensitive to changes in IKK concentration.
We have shown that the spiky oscillations 
can, indeed, show a high sensitivity to IKK.
This sensitivity 
allows a great versatility in the regulation of downstream genes
by NF-$\kappa$B. Where the cell requires a gene to be very sensitive
to the IKK concentration, the NF-$\kappa$B system can result in steep response
curves with Hill coefficients larger than 20. And where a slower response
is necessary, it can be achieved by adjusting the
binding and dissociation constants of NF-$\kappa$B to that operator site.
Further, we found that 
cascades of different length could be used to turn certain
genes on earlier or later. Given this versatility in regulatory strategies
it seems likely that
cells would have evolved to make use of these properties of the
NF-$\kappa$B oscillations.
It remains for future experiments to uncover the particular ways
NF-$\kappa$B regulates specific genes.\\

\noindent
We thank J. Ferkinghoff-Borg, E. Siggia and G. Tiana for useful discussions.
This work was supported by the Danish Research Foundation.

\bibliography{nfkbosc}

\clearpage
\onecolumn
\section*{Supporting Text}
\subsection*{7-variable model of NF-$\kappa$B signalling}
We use the following abbreviations:
$N_n \& N$, free nuclear and cytoplasmic
NF-$\kappa$B; 
$I_m$, I$\kappa$B mRNA; $I_n \& I$, free nuclear and cytoplasmic I$\kappa$B; 
$(NI)_n \& (NI)$, nuclear and cytoplasmic NF-$\kappa$B--I$\kappa$B complex;
IKK, I$\kappa$B kinase.
The 7-variable model is defined by the equations: 
$$\frac{dN_n}{dt}=k_{Nin}N-k_{fn}N_nI_n+k_{bn}(NI)_n,$$
$$\frac{dI_m}{dt}=k_tN_n^2-\gamma_mI_m,$$
$$\frac{dI}{dt}=k_{tl}I_m-k_fNI+k_b(NI)-k_{Iin}I+k_{Iout}I_n,$$
$$\frac{dN}{dt}=-k_fNI+(k_b+\alpha)(NI)-k_{Nin}N,$$
$$\frac{d(NI)}{dt}=k_fNI-(k_b+\alpha)(NI)+k_{NIout}(NI)_n,$$
$$\frac{dI_n}{dt}=k_{Iin}I-k_{Iout}I_n-k_{fn}N_nI_n+k_{bn}(NI)_n,$$
$$\frac{d(NI)_n}{dt}=k_{fn}N_nI_n-(k_{bn}+k_{NIout})(NI)_n.$$

Figure \ref{suppreducfig} shows a plot of nuclear NF-$\kappa$B concentration 
obtained by integrating these equations using the following
parameter values (Hoffmann et al. (2002) {\it Science} {\bf 298}, 1241):
$k_{Nin}=5.4~{\rm min}^{-1},k_{Iin}=0.018~{\rm min}^{-1},k_{Iout}=0.012~{\rm min}^{-1},k_{NIout}=0.83~{\rm min}^{-1},k_t=1.03~\mu M^{-1}{\rm min}^{-1},k_{tl}=0.24~{\rm min}^{-1},
k_f=k_{fn}=30~\mu M^{-1}{\rm min}^{-1},k_b=k_{bn}=0.03~{\rm min}^{-1},\alpha=1.05\times{\rm IKK}~{\rm min}^{-1},\gamma_m=0.017 {\rm min}^{-1}$. 
The initial conditions were $N=1 \mu M$, ${\rm IKK}=0.5 \mu M$ and all other concentrations zero.

\subsection*{Reduction from 7-variables to 3-variables}
First, taking note of the fact that $k_f$ and $k_{fn}$ are large, we assume that
all complexes are in equilibrium, i.e.:
$$k_fNI\approx(k_b+\alpha)(NI),$$
$$k_{fn}N_nI_n\approx(k_{bn}+k_{NIout})(NI)_n.$$
Simulations show that these are good approximations.
In terms of $I_n^{tot}\equiv I_n+(NI)_n$ and $N_c^{tot}\equiv N+(NI)=N_{tot}-N_n$, which are slowly varying,
we can rewrite the above equations as follows:
$$(NI)=(N_{tot}-N_n)\frac{I}{K_I+I},$$
$$N=(N_{tot}-N_n)\frac{K_I}{K_I+I},$$
$$(NI)_n=I_n^{tot}\frac{N_n}{K_N+N_n},$$
$$I_n=I_n^{tot}\frac{K_N}{K_N+N_n},$$
where $K_I\equiv(k_b+\alpha)/k_f=0.035 \mu M$ and $K_N\equiv(k_{bn}+k_{NIout})/k_{fn}=0.029 \mu M$, using the parameter values above.

Using these expressions, the equations of the 7-variable model reduce to the following four (Fig. \ref{suppreducfig}):
$$\frac{dN_n}{dt}=k_{Nin}K_I\frac{(N_{tot}-N_n)}{K_I+I}-k_{NIout}\frac{I_n^{tot} N_n}{K_N+N_n},$$
$$\frac{dI_m}{dt}=k_tN_n^2-\gamma_mI_m,$$
$$\frac{dI}{dt}=k_{tl}I_m-\alpha \frac{(N_{tot}-N_n)I}{K_I+I}-k_{Iin}I+k_{Iout}K_N\frac{I_n^{tot}}{K_N+N_n},$$
$$\frac{dI_n^{tot}}{dt}=k_{Iin}I-k_{Iout}K_N\frac{I_n^{tot}}{K_N+N_n}-k_{NIout}\frac{I_n^{tot}N_n}{K_N+N_n}.$$

First, we note that the terms $-k_{Iin}I$ and $k_{Iout}K_N\frac{I_n^{tot}}{K_N+N_n}$
in the $dI/dt$ equation
are much smaller than $-\alpha \frac{(N_{tot}-N_n)I}{K_I+I}$ and can be neglected
as long as IKK is non-zero. Secondly, simulations reveal that the term 
$k_{NIout}\frac{I_n^{tot} N_n}{K_N+N_n}$, in the $dI_n^{tot}/dt$ equation, also shows sharp spikes as a function of time
which coincide with the spikes of $N_n$. The value of this term is substantial
only when $N_n\gg K_N$, i.e., during the spikes of $N_n$, and at those times $I_n^{tot}$ dips to its minimum.
We therefore make the approximation that $I_n^{tot}$ can be
replaced by its minimum value, $I_{n,min}^{tot}$, which satisfies the equation:
$$k_{Iin}I=k_{Iout}K_N\frac{I_{n,min}^{tot}}{K_N+N_n}+k_{NIout}\frac{I_{n,min}^{tot}N_n}{K_N+N_n}.$$
In the regime where $N_n\gg K_n$ this gives
$$I_{n,min}^{tot}\approx \frac{k_{Iin}}{k_{NIout}}I.$$

Using this we can reduce to a 3-variable model:
$$\frac{dN_n}{dt}=k_{Nin}K_I\frac{(N_{tot}-N_n)}{K_I+I}-k_{Iin}\frac{IN_n}{\delta+N_n},$$
$$\frac{dI_m}{dt}=k_tN_n^2-\gamma_mI_m,$$
$$\frac{dI}{dt}=k_{tl}I_m-\alpha\frac{(N_{tot}-N_n)I}{K_I+I}.$$

\renewcommand{\thefigure}{S\arabic{figure}}
\setcounter{figure}{0}
\begin{figure}
\input{reductionfig.pstex_t}
\caption{\label{suppreducfig}Schematic showing reduction of the 7-variable to a 3-variable model, as well as plots of oscillatory
nuclear NF-$\kappa$B from simulations of the 7 and 3-variable models using parameter values given in the text.
}
\end{figure}
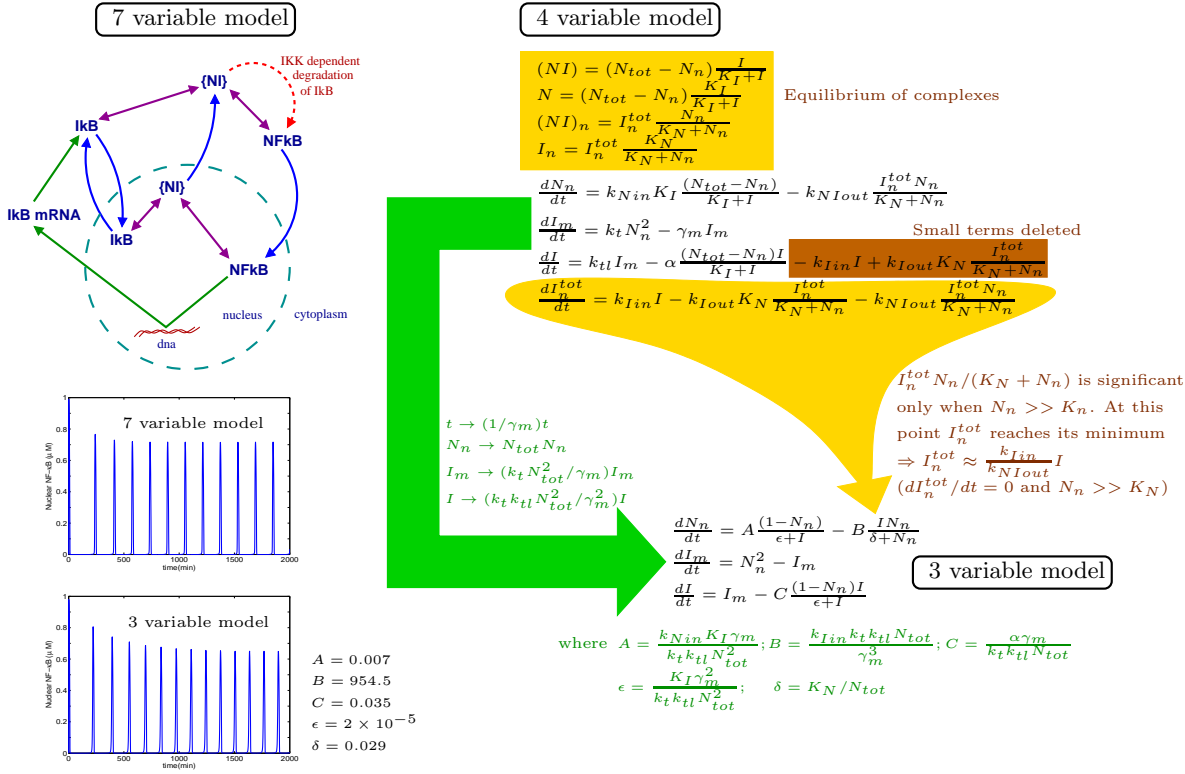

\subsection*{Rescaling the 3-variable model}
For ease of analysis, we reduce the number of parameters in the model by rescaling
all variables to become dimensionless.
We use the following transformations on the above equations:
$$t\rightarrow (1/\gamma_m)t,$$
$$N_n\rightarrow N_{tot}N_n,$$
$$I_m\rightarrow (k_tN_{tot}^2/\gamma_m)I_m,$$
$$I\rightarrow (k_tk_{tl}N_{tot}^2/\gamma_m^2)I,$$
which gives:
$$\frac{dN_n}{dt}=A\frac{(1-N_n)}{\epsilon+I}-B\frac{IN_n}{\delta+N_n},$$
$$\frac{dI_m}{dt}=N_n^2-I_m$$
$$\frac{dI~~}{dt}=I_m-C\frac{(1-N_n)I}{\epsilon+I},$$
with
$$A=\frac{k_{Nin}K_I\gamma_m}{k_tk_{tl}N_{tot}^2}\approx 0.007,$$
$$B=\frac{k_{Iin}k_tk_{tl}N_{tot}}{\gamma_m^3}\approx 954.5,$$
$$C=\frac{\alpha\gamma_m}{k_tk_{tl}N_{tot}}\approx 0.035,$$
$$\delta=\frac{K_N}{N_{tot}}\approx 0.029,$$
$$\epsilon=\frac{K_I\gamma_m^2}{k_tk_{tl}N_{tot}^2}\approx 2\times 10^{-5}.$$

\subsection*{Steady state solution of the 3-variable model}
The steady state values of $N_n$, $I_m$ and $I$ are solutions to
$$A\frac{(1-N_n)}{\epsilon+I}-B\frac{IN_n}{\delta+N_n}=0,$$
$$N_n^2-I_m=0,$$
$$I_m-C\frac{(1-N_n)I}{\epsilon+I}=0.$$

$I_m$ and $I$ can be eliminated
using 
$$I_m=N_n^2,$$
$$I=\frac{N_n^2\epsilon}{C-CN_n-N_n^2}.$$

From this we find that the steady state value of $N_n$ is a solution of the equation
$$(C-CN_n-N_n^2)^2=\frac{BC\epsilon^2}{A}\frac{N_n^3}{\delta+N_n},$$
or equivalently,
$$N_n^5+(\delta+2C)N_n^4+C\left[2(\delta-1)+C-\frac{B}{A}\epsilon^2\right]N_n^3+C[(C-2)\delta-2C]N_n^2+C^2(1-2\delta)N_n+C^2\delta=0.$$

In general, this has two real solutions, one with $C-CN_n-N_n^2>0$ and the other with
$C-CN_n-N_n^2<0$. The latter results in a negative value for $I$ and therefore is not
an acceptable solution. Thus we are left with only one fixed point.

\subsection*{Linear stability of the fixed point}
We linearize the equations around the fixed point, which gives the Hessian
$$J=\left(
\begin{array}{ccc}
-\frac{A}{\epsilon+I}-\frac{\delta BI}{(\delta+N_n)^2} & 0 & -\frac{A(1-N_n)}{(\epsilon+I)^2}-\frac{BN_n}{\delta+N_n} \\
2N_n & -1 & 0\\
\frac{CI}{\epsilon+I} & 1 & -\frac{C\epsilon(1-N_n)}{(\epsilon+I)^2}\\
\end{array}
\right)$$

This matrix can be used to examine the stability of the fixed point. 
If $\lambda_i$ are the (possibly complex) eigenvalues of this matrix, then 
the fixed point is unstable if
$${\rm max}_i \left[{\rm Re}(\lambda_i)\right]>0$$
and stable if 
$${\rm max}_i \left[{\rm Re}(\lambda_i)\right]<0.$$

\subsection*{Linear production of $I_m$}
In this section we consider the effect of taking the production of $I_m$ to
be linear in $N_n$, instead of $N_n^2$:
$$\frac{dN_n}{dt}=k_{Nin}K_I\frac{(N_{tot}-N_n)}{K_I+I}-k_{Iin}\frac{IN_n}{\delta+N_n},$$
$$\frac{dI_m}{dt}=k_tN_n-\gamma_mI_m,$$
$$\frac{dI}{dt}=k_{tl}I_m-\alpha\frac{(N_{tot}-N_n)I}{K_I+I}.$$

Using very similar transformations, we rescale the variables:
$$t\rightarrow (1/\gamma_m)t,$$
$$N_n\rightarrow N_{tot}N_n,$$
$$I_m\rightarrow (k_tN_{tot}/\gamma_m)I_m,$$
$$I\rightarrow (k_tk_{tl}N_{tot}/\gamma_m^2)I,$$
which gives:
$$\frac{dN_n}{dt}=A\frac{(1-N_n)}{\epsilon+I}-B\frac{IN_n}{\delta+N_n},$$
$$\frac{dI_m}{dt}=N_n-I_m,$$
$$\frac{dI~~}{dt}=I_m-C\frac{(1-N_n)I}{\epsilon+I},$$
with
$$A=\frac{k_{Nin}K_I\gamma_m}{k_tk_{tl}N_{tot}},$$
$$B=\frac{k_{Iin}k_tk_{tl}}{\gamma_m^3},$$
$$C=\frac{\alpha\gamma_m}{k_tk_{tl}},$$
$$\delta=\frac{K_N}{N_{tot}},$$
$$\epsilon=\frac{K_I\gamma_m^2}{k_tk_{tl}N_{tot}}.$$

\begin{figure}
{\bf A.\hfill B.\hfill ~}\\
\epsfig{height=5cm,file=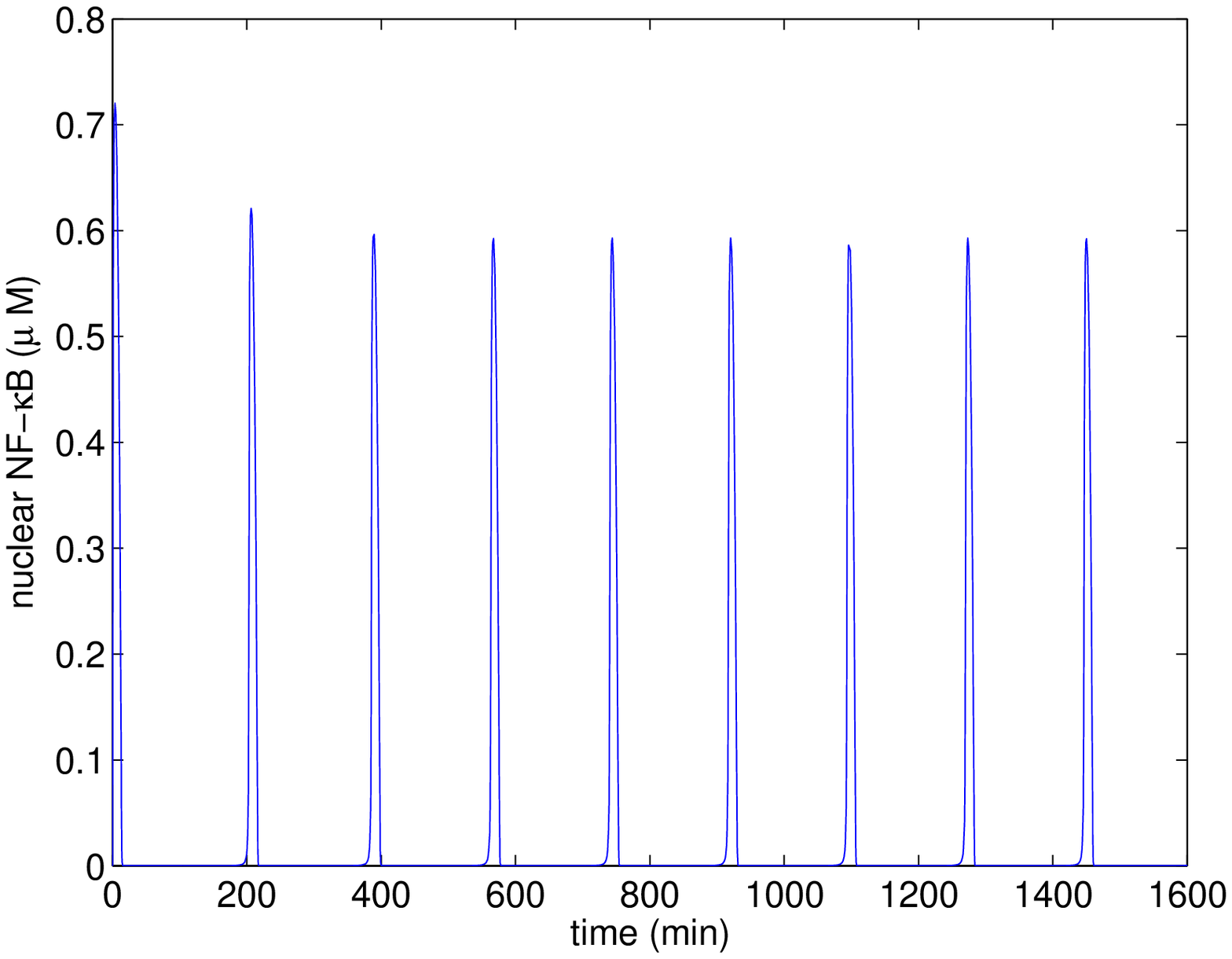}\hfill
\epsfig{height=5cm,file=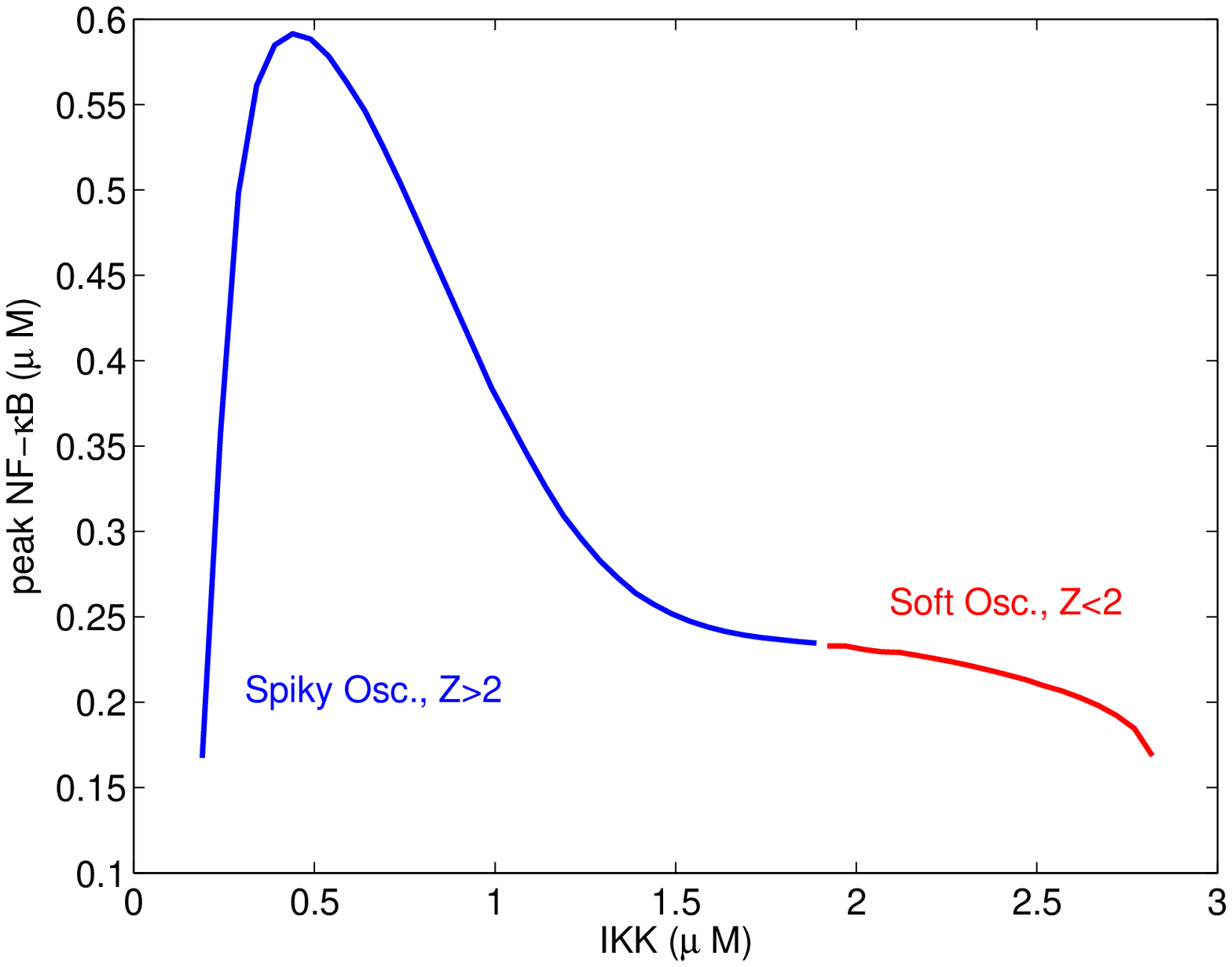}\hfill ~\\
\caption{\label{linearIm_osc}
{\bf A.} Sustained spiky oscillations of nuclear NF-$\kappa$B when the production of $I_m$ is linearly dependent
on $N_n$. Parameter values: $A=0.001$, $B=954.5$, $C=0.035$, $\delta=0.029$ and
$\epsilon=2\times 10^{-5}$.
{\bf B.} Spike peak, the maximum concentration of nuclear NF-$\kappa$B, as a function of IKK concentration. 
Blue and red signify, respectively, regions of spiky
and soft oscillations. 
}
\end{figure}

If we use the same parameter values as before we do not get sustained
oscillations. Not surprisingly, the region in parameter space
where we get sustained oscillations has shifted. Simply taking
$A=0.001$, we get spiky oscillations, as in the original model (Fig. \ref{linearIm_osc}A).
Fig. \ref{linearIm_osc}B shows the response to changes in IKK.

The steady state value of $N_n$ is now a solution to
$$(C-CN_n-N_n)^2=\frac{BC\epsilon^2}{A}\frac{N_n^2}{\delta+N_n}.$$
The steady value of $I$ can be calculated from the value of $N_n$ using:
$$I=\frac{N_n\epsilon}{C-CN_n-N_n}.$$

In general, there are again two real solutions, one with $C-CN_n-N_n>0$ and the other with
$C-CN_n-N_n<0$. The latter results in a negative value for $I$ and therefore is not
an acceptable solution. Thus we are left with only one fixed point.

\begin{figure}
\centerline{\epsfig{height=5cm,file=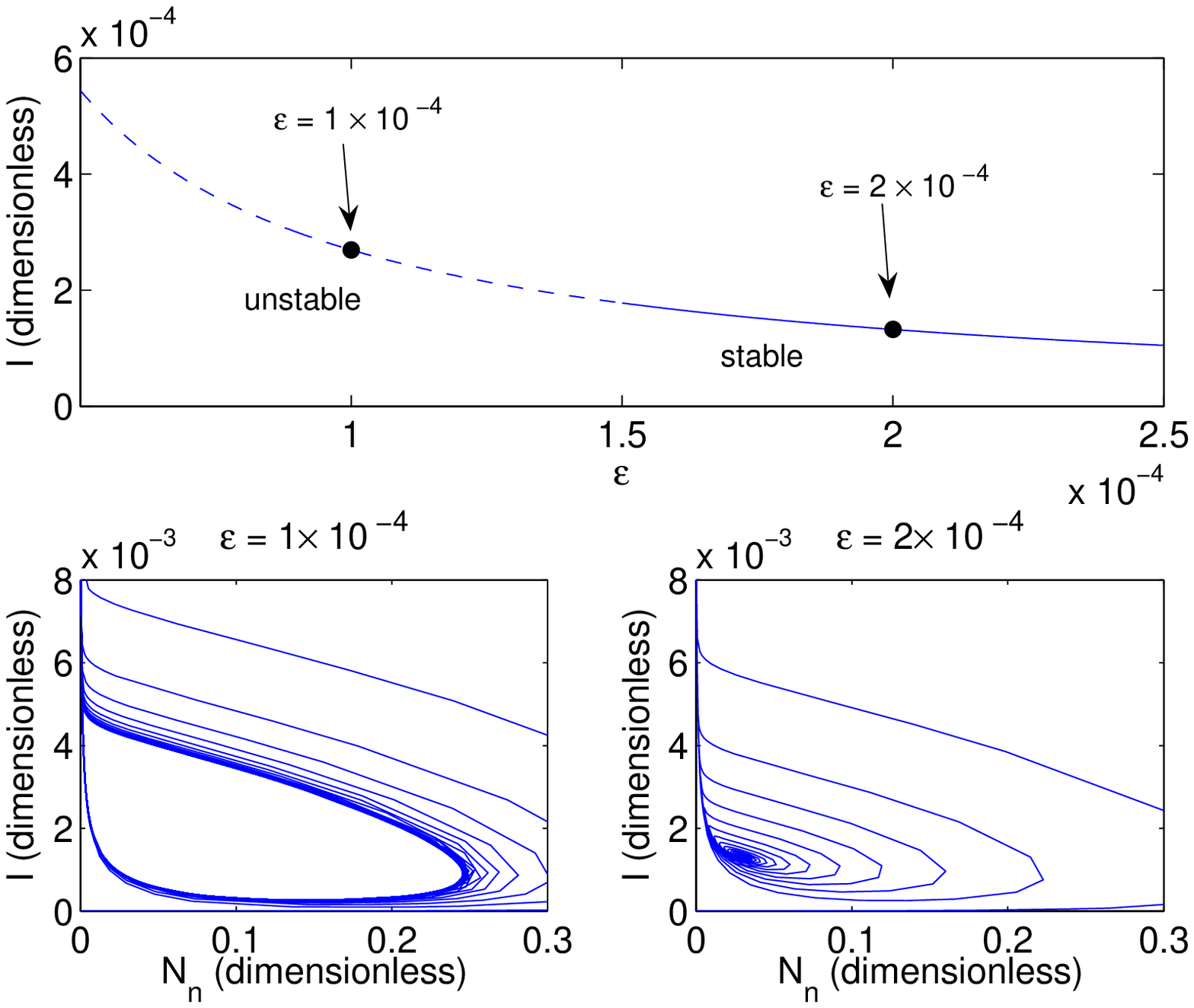}}
\caption{\label{linearIm_stability}
{\bf A.} Steady state solution for $I$ as a function of the parameter $\epsilon$,
when the production of $I_m$ is linearly dependent on $N_n$.
Dashed line shows the region where the fixed point is unstable and the solid line
shows the region where it is stable. Black dots mark the parameter values used in (B) and (C).
The crossover from unstable to stable occurs where $\epsilon$ becomes comparable to the steady
state value of $I$, i.e., where the degradation of $I$ stops being saturated.
{\bf B.} Trajectory of the system in the $I$-$N_n$ phase plane, for $\epsilon=1\times 10^{-4}$, which converges to
a stable limit cycle. {\bf C.} Trajectory of the system, for $\epsilon=2\times 10^{-4}$, which converges
to a stable fixed point.
}
\end{figure}

Linearizing the equations around the fixed point gives almost the same Hessian. The only difference
is in one matrix element: $J_{2,1}=1$ instead of $2N_n$.
Fig. \ref{linearIm_stability} shows the stability of the fixed point as a function of $\epsilon$.

\end{document}

%% file: reductionfig.pstex_t
\begin{picture}(0,0)%
\includegraphics{reductionfig.pstex}%
\end{picture}%
\setlength{\unitlength}{1776sp}%
\begingroup\makeatletter\ifx\SetFigFont\undefined%
\gdef\SetFigFont#1#2#3#4#5{%
  \reset@font\fontsize{#1}{#2pt}%
  \fontfamily{#3}\fontseries{#4}\fontshape{#5}%
  \selectfont}%
\fi\endgroup%
\begin{picture}(15447,10758)(187,-9286)
\put(7576,-1786){\makebox(0,0)[lb]{\smash{\SetFigFont{6}{7.2}{\familydefault}{\mddefault}{\updefault}{\color[rgb]{0,0,0}$\frac{dI_m}{dt}=k_tN_n^2-\gamma_mI_m$}%
}}}
\put(7576,-2761){\makebox(0,0)[lb]{\smash{\SetFigFont{6}{7.2}{\familydefault}{\mddefault}{\updefault}{\color[rgb]{0,0,0}$\frac{dI_n^{tot}}{dt}=k_{Iin}I-k_{Iout}K_N\frac{I_n^{tot}}{K_N+N_n}-k_{NIout}\frac{I_n^{tot}N_n}{K_N+N_n}$}%
}}}
\put(7876,-7561){\makebox(0,0)[lb]{\smash{\SetFigFont{6}{7.2}{\rmdefault}{\mddefault}{\updefault}{\color[rgb]{0,.56,0}where}%
}}}
\put(7576,-2236){\makebox(0,0)[lb]{\smash{\SetFigFont{6}{7.2}{\familydefault}{\mddefault}{\updefault}{\color[rgb]{0,0,0}$\frac{dI}{dt}=k_{tl}I_m-\alpha \frac{(N_{tot}-N_n)I}{K_I+I}-k_{Iin}I+k_{Iout}K_N\frac{I_n^{tot}}{K_N+N_n}$}%
}}}
\put(7576,-1261){\makebox(0,0)[lb]{\smash{\SetFigFont{6}{7.2}{\familydefault}{\mddefault}{\updefault}{\color[rgb]{0,0,0}$\frac{dN_n}{dt}=k_{Nin}K_I\frac{(N_{tot}-N_n)}{K_I+I}-k_{NIout}\frac{I_n^{tot} N_n}{K_N+N_n}$}%
}}}
\put(9451,-6436){\makebox(0,0)[lb]{\smash{\SetFigFont{6}{7.2}{\familydefault}{\mddefault}{\updefault}{\color[rgb]{0,0,0}$\frac{dI_m}{dt}=N_n^2-I_m$}%
}}}
\put(9451,-6886){\makebox(0,0)[lb]{\smash{\SetFigFont{6}{7.2}{\familydefault}{\mddefault}{\updefault}{\color[rgb]{0,0,0}$\frac{dI}{dt}=I_m-C\frac{(1-N_n)I}{\epsilon+I}$}%
}}}
\put(7576,464){\makebox(0,0)[lb]{\smash{\SetFigFont{6}{7.2}{\familydefault}{\mddefault}{\updefault}{\color[rgb]{0,0,0}$(NI)=(N_{tot}-N_n)\frac{I}{K_I+I}$}%
}}}
\put(7576, 89){\makebox(0,0)[lb]{\smash{\SetFigFont{6}{7.2}{\familydefault}{\mddefault}{\updefault}{\color[rgb]{0,0,0}$N=(N_{tot}-N_n)\frac{K_I}{K_I+I}$}%
}}}
\put(7576,-286){\makebox(0,0)[lb]{\smash{\SetFigFont{6}{7.2}{\familydefault}{\mddefault}{\updefault}{\color[rgb]{0,0,0}$(NI)_n=I_n^{tot}\frac{N_n}{K_N+N_n}$}%
}}}
\put(7576,-661){\makebox(0,0)[lb]{\smash{\SetFigFont{6}{7.2}{\familydefault}{\mddefault}{\updefault}{\color[rgb]{0,0,0}$I_n=I_n^{tot}\frac{K_N}{K_N+N_n}$}%
}}}
\put(9451,-5986){\makebox(0,0)[lb]{\smash{\SetFigFont{6}{7.2}{\familydefault}{\mddefault}{\updefault}{\color[rgb]{0,0,0}$\frac{dN_n}{dt}=A\frac{(1-N_n)}{\epsilon+I}-B\frac{IN_n}{\delta+N_n}$}%
}}}
\put(10801,-7561){\makebox(0,0)[lb]{\smash{\SetFigFont{5}{6.0}{\familydefault}{\mddefault}{\updefault}{\color[rgb]{0,.56,0}$B=\frac{k_{Iin}k_tk_{tl}N_{tot}}{\gamma_m^3}$;}%
}}}
\put(6301,-5536){\makebox(0,0)[lb]{\smash{\SetFigFont{5}{6.0}{\familydefault}{\mddefault}{\updefault}{\color[rgb]{0,.56,0}$I\rightarrow (k_tk_{tl}N_{tot}^2/\gamma_m^2)I$}%
}}}
\put(6301,-5161){\makebox(0,0)[lb]{\smash{\SetFigFont{5}{6.0}{\familydefault}{\mddefault}{\updefault}{\color[rgb]{0,.56,0}$I_m\rightarrow (k_tN_{tot}^2/\gamma_m)I_m$}%
}}}
\put(1801,-4486){\makebox(0,0)[lb]{\smash{\SetFigFont{6}{7.2}{\rmdefault}{\mddefault}{\updefault}{\color[rgb]{0,0,0}7 variable model}%
}}}
\put(1876,-7261){\makebox(0,0)[lb]{\smash{\SetFigFont{6}{7.2}{\rmdefault}{\mddefault}{\updefault}{\color[rgb]{0,0,0}3 variable model}%
}}}
\put(13051,-6586){\makebox(0,0)[lb]{\smash{\SetFigFont{9}{10.8}{\familydefault}{\mddefault}{\updefault}{\color[rgb]{0,0,0}3 variable model}%
}}}
\put(7576,1139){\makebox(0,0)[lb]{\smash{\SetFigFont{9}{10.8}{\familydefault}{\mddefault}{\updefault}{\color[rgb]{0,0,0}4 variable model}%
}}}
\put(10876,-8161){\makebox(0,0)[lb]{\smash{\SetFigFont{5}{6.0}{\familydefault}{\mddefault}{\updefault}{\color[rgb]{0,.56,0}$\delta=K_N/N_{tot}$}%
}}}
\put(6301,-4786){\makebox(0,0)[lb]{\smash{\SetFigFont{5}{6.0}{\familydefault}{\mddefault}{\updefault}{\color[rgb]{0,.56,0}$N_n\rightarrow N_{tot}N_n$}%
}}}
\put(6301,-4486){\makebox(0,0)[lb]{\smash{\SetFigFont{5}{6.0}{\familydefault}{\mddefault}{\updefault}{\color[rgb]{0,.56,0}$t\rightarrow (1/\gamma_m)t$}%
}}}
\put(4426,-7786){\makebox(0,0)[lb]{\smash{\SetFigFont{5}{6.0}{\familydefault}{\mddefault}{\updefault}{\color[rgb]{0,0,0}$A=0.007$}%
}}}
\put(4426,-8086){\makebox(0,0)[lb]{\smash{\SetFigFont{5}{6.0}{\familydefault}{\mddefault}{\updefault}{\color[rgb]{0,0,0}$B=954.5$}%
}}}
\put(4426,-8386){\makebox(0,0)[lb]{\smash{\SetFigFont{5}{6.0}{\familydefault}{\mddefault}{\updefault}{\color[rgb]{0,0,0}$C=0.035$}%
}}}
\put(4426,-8686){\makebox(0,0)[lb]{\smash{\SetFigFont{5}{6.0}{\familydefault}{\mddefault}{\updefault}{\color[rgb]{0,0,0}$\epsilon=2\times 10^{-5}$}%
}}}
\put(4426,-8986){\makebox(0,0)[lb]{\smash{\SetFigFont{5}{6.0}{\familydefault}{\mddefault}{\updefault}{\color[rgb]{0,0,0}$\delta=0.029$}%
}}}
\put(1651,1139){\makebox(0,0)[lb]{\smash{\SetFigFont{9}{10.8}{\familydefault}{\mddefault}{\updefault}{\color[rgb]{0,0,0}7 variable model}%
}}}
\put(12601,-3886){\makebox(0,0)[lb]{\smash{\SetFigFont{6}{7.2}{\familydefault}{\mddefault}{\updefault}{\color[rgb]{.5,.17,0}$I_n^{tot}N_n/(K_N+N_n)$ is significant}%
}}}
\put(12601,-4261){\makebox(0,0)[lb]{\smash{\SetFigFont{6}{7.2}{\familydefault}{\mddefault}{\updefault}{\color[rgb]{.5,.17,0}only when $N_n>>K_n$. At this}%
}}}
\put(12601,-4636){\makebox(0,0)[lb]{\smash{\SetFigFont{6}{7.2}{\familydefault}{\mddefault}{\updefault}{\color[rgb]{.5,.17,0}point $I_n^{tot}$ reaches its minimum}%
}}}
\put(12601,-5011){\makebox(0,0)[lb]{\smash{\SetFigFont{6}{7.2}{\familydefault}{\mddefault}{\updefault}{\color[rgb]{.5,.17,0}$\Rightarrow I_n^{tot}\approx\frac{k_{Iin}}{k_{NIout}}I$}%
}}}
\put(12601,-5386){\makebox(0,0)[lb]{\smash{\SetFigFont{6}{7.2}{\familydefault}{\mddefault}{\updefault}{\color[rgb]{.5,.17,0}($dI_n^{tot}/dt=0$ and $N_n>>K_N$)}%
}}}
\put(8701,-7561){\makebox(0,0)[lb]{\smash{\SetFigFont{5}{6.0}{\familydefault}{\mddefault}{\updefault}{\color[rgb]{0,.56,0}$A=\frac{k_{Nin}K_I\gamma_m}{k_tk_{tl}N_{tot}^2}$;}%
}}}
\put(8701,-8161){\makebox(0,0)[lb]{\smash{\SetFigFont{5}{6.0}{\familydefault}{\mddefault}{\updefault}{\color[rgb]{0,.56,0}$\epsilon=\frac{K_I\gamma_m^2}{k_tk_{tl}N_{tot}^2}$;}%
}}}
\put(11026, 89){\makebox(0,0)[lb]{\smash{\SetFigFont{6}{7.2}{\rmdefault}{\mddefault}{\updefault}{\color[rgb]{.5,.17,0}Equilibrium of complexes}%
}}}
\put(12826,-1786){\makebox(0,0)[lb]{\smash{\SetFigFont{6}{7.2}{\rmdefault}{\mddefault}{\updefault}{\color[rgb]{.5,.17,0}Small terms deleted}%
}}}
\put(13276,-7561){\makebox(0,0)[lb]{\smash{\SetFigFont{5}{6.0}{\familydefault}{\mddefault}{\updefault}{\color[rgb]{0,.56,0}$C=\frac{\alpha\gamma_m}{k_tk_{tl}N_{tot}}$}%
}}}
\end{picture}